\documentclass[10pt]{article}

\usepackage[english]{babel}
\usepackage{color}
\usepackage{amsmath}
\usepackage{amssymb}
\usepackage{amsfonts}
\usepackage{amsthm}
\usepackage{amstext}
\usepackage{alltt}
\usepackage{verbatim}
\usepackage{array}
\usepackage{booktabs}
\usepackage[colorlinks=true,linkcolor=blue]{hyperref}
\usepackage{graphicx}
\usepackage{float}
\usepackage[lined,boxed]{algorithm2e}
\usepackage{authblk}
\usepackage{nicefrac}

\usepackage{geometry}
\newcommand{\ind}{\stackrel{\mathrm{ind.}}{\sim}}

\title{Relative Frequencies of Constrained Events in\\ Stochastic Processes: an Analytical Approach}

\author[1]{S. Rusconi \thanks{srusconi@bcamath.org}}
\author[1,2]{E. Akhmatskaya}
\author[2,3]{D. Sokolovski}
\author[4]{N. Ballard}
\author[4]{J.C. de la Cal}

\affil[1]{BCAM - Basque Center for Applied Mathematics, Alameda de Mazarredo 14, 48009 Bilbao, Bizkaia, Spain}
\affil[2]{IKERBASQUE, Basque Foundation for Science, E-48013 Bilbao, Spain}
\affil[3]{Departmento de Qu\'imica-F\'isica, Universidad del Pa\' is Vasco, UPV/EHU, Leioa, Spain}
\affil[4]{POLYMAT and Grupo de Ingenier\'ia Qu\'imica, Dpto. de Qu\'mica Aplicada, University of the Basque Country UPV/EHU, Joxe Mari Korta Zentroa, Tolosa Etorbidea 72, 20018 Donostia/San Sebasti\'an, Spain}

\date{}

\begin{document}

\maketitle

\begin{abstract}
Stochastic Simulation Algorithm (SSA) and the corresponding Monte Carlo (MC) method are ones of the most common approaches for studying stochastic processes. They rely on the knowledge of inter-events probability density functions (pdf),  and on the information about dependencies between all possible events. 
Analytical representations of a pdf are difficult to specify in advance, in many real life applications. 
Knowing a shape of pdf, and using experimental data, different optimization schemes can be applied in order to evaluate the probability density functions and, therefore, the properties of the studied system. Such methods, however, are computationally demanding, and often not feasible. We show that in the case where experimentally accessed properties are directly related to the frequencies of events involved, it may be possible to replace the heavy Monte Carlo core of optimization schemes with an analytical solution. Such replacement not only provides more accurate estimation of the properties of the process, but also reduces the simulation time by a factor of order of the sample size (at least $\approx 10^4$). The proposed analytical approach is valid for any choice of the pdf. The accuracy, computational efficiency, and advantages of the method over MC procedures, are demonstrated in the exactly solvable case and in evaluation of branching fractions in Controlled Radical Polymerization (CRP) of acrylic monomers. This polymerization can be modelled by a constrained stochastic process. Constrained systems are quite common, and this makes the method useful for various applications. 
\end{abstract}

\section{Introduction}
\label{sec:intro}

	\noindent Simulation of stochastic processes is a powerful tool for modelling and describing the evolution of various phenomena in natural and human-made systems.
	 A well known example of such modelling is the Stochastic Simulation Algorithm (SSA) developed by D. Gillespie \cite{GillespieOriginalSSA}. In physics community, it is also known as the \emph{n-fold way}, introduced by Bortz et al. \cite{Bortz1975}. SSA is a Monte Carlo (MC) based method: it draws multiple realizations of the process and then computes statistics on them. From now on, we will refer to SSA as MC method. This approach is based on the assumption that the studied system is well-mixed, and also memoryless. These assumptions lead to independent exponentially distributed inter-event times probability density functions \cite{GillespieReviewSSAminimumTime}.
	  As intuition suggests, this set of hypothesis does not hold for all phenomena of practical interest. One such example is a constrained stochastic process, where the occurrence of some events may depend on the previous history of the process (see \cite{ConstrainedSystemExamples} for examples of constrained stochastic processes). In this cases, the dependencies can be realized either {\it explicitly} by introducing constraints in the SSA algorithm, or {\it implicitly}, through a modification of inter-event times probability density functions (pdf).\\
	\noindent Either way, well posed formulations of stochastic processes require the knowledge of inter-event times probability density functions. One can assume a particular functional shape for these pdf, but meaningful values of the pdf parameters are also needed, in order to complete the description of the model and run the corresponding Monte Carlo simulation. In practice, appropriate pdf parameters are difficult to identify.\\
	\noindent One possible way to estimate the unknown parameters is to employ an optimisation scheme, which uses available experimental data. The idea behind a fitting scheme is to build a cost function $J$, which measures the disagreement between the experimental data, and the data obtained by the proposed MC method. A fitting scheme seeks to minimize the cost function $J$, in order to find the set of pdf parameters which gives the best agreement with the experimental data.\\
	\noindent Since the stochastic simulation is a part of the cost function evaluation, multiple runs are needed until a good fitting is obtained. Thus, regardless of the choice of the optimization route, and of the particular cost function, any MC based fitting scheme is computationally expensive. 
	
	\noindent In this paper we derive an analytical approach for evaluation of $J$, as a function of pdf's parameters, without running the corresponding MC simulation.
	 The method is free of the statistical errors affecting Monte Carlo based simulations. 
	 It can be incorporated into fitting schemes used to study various phenomena for which the experimentally observed properties are directly related to frequencies of the events involved. Examples of such phenomena include  branching rate reduction in Controlled Radical Polymerization \cite{Ahmad_et_al}, high-frequency pulsed laser polymerization of acrylates \cite{ReyesArzamendi2011}, reduction in rate of polymerization in RAFT-mediated polymerizations \cite{Barner-Kowollik2006,Moad}, and the absence of chain transfer to PVA in SET-DTLRP of PVC in aqueous media \cite{Coelho}.\\
	\noindent Our goal is to provide analytical expressions for the asymptotic frequencies of events involved in constrained stochastic processes, thus avoiding costly Monte Carlo evaluations, used in a fitting scheme. 
	A natural application for the new method is the computation of branching fractions in Controlled Radical Polymerization (CRP) of acrylic monomers \cite{CRPexplaned}.\\	
	\noindent The paper is organized as follows. In \autoref{sec:generic_asymptotic_fractions} we propose a general approach for calculation of asymptotic relative frequencies of events in constrained stochastic processes. 
	It is solved analytically in \autoref{sec:one_constraint} for a process with a single constraint, and the results are compared with the ones earlier provided in \cite{TimeDelayedPDF}. 
	CRP and the formation of branches in CRP of acrylates is briefly described in \autoref{sec:CRP}.
	 In \autoref{sec:MCfitting} the MC fitting method earlier proposed for evaluation of the branching fraction in the CRP \cite{LinExpPdf} is summarized. 
	 Our alternative analytical approach is developed in \autoref{sec:anal_sol}. The results of comparison between two different methodologies are provided in  \autoref{sec:results}. 
	 The concluding remarks are given in \autoref{sec:sum}.
		
\section{Relative Asymptotic Frequencies of Events in Constrained\\ Stochastic Processes}
\label{sec:generic_asymptotic_fractions}

	\noindent Consider a stochastic process with a given total number of events $n_T >> 1$. These events correspond to the realizations of the random variables building the stochastic process. Each event may have $N$ possible outcomes. A particular outcome is given by a possible value assumed by the random variables. The outcomes are constrained, i.e. the outcome $i=1,..,N$ can occur if and only if at least $c_{ij}$ outcomes $j=1,..,N$ have already occurred after the previous occurrence of the $i$-th outcome.\\
	\noindent Our objective is to compute ratios between asymptotic numbers of occurred outcomes, $n_i$ and $n_j$, corresponding to two events of different types, $i$ and $j$.\\ 
	 \noindent It is worthy to remark that there are the conditions the system needs to satisfy in order to be able to evolve its state. First, each outcome must not be constrained by itself, or
 	
	\begin{equation} 	
 	c_{ii} = 0 \quad \forall i= \{1,..,N \},
 	\end{equation}
 	
 	\noindent and second, there is at least one outcome free to occur, or (s.t. is a shorthand for ``such that'')	
 	
	\begin{equation} 	
 	\exists \,  i \in \{1,..,N\} \quad \mbox{s.t.} \quad c_{ij} = 0 \, \forall j=\{1,..,N\}.
 	\end{equation} 	
 	
	\noindent Otherwise, if $c_{ij} \neq 0 \land c_{ji} \neq 0$ for all $i$, $j$, none of events are possible.\\
	\noindent The $n_T$ events can be partitioned as follows. Let us divide all events into $2^N$ non-overlapping subsets labelled $(j_1, j_ 2, .. , j_N)$, where $j_k = 0,1$ and $k=1,2,..,N$. $j_k=1$ means that the $k$-th outcome is possible, whereas if $j_k=0$ it is not possible due to constraints. 
	Let $n(j_1,j_2,..,j_N)$ be the number of events in a subset, e.g.,

	\begin{equation}
	n(1, 0, ..., 0)
	\end{equation}
	
	\noindent is the number of events for which the outcome $i=1$ is possible, but the rest of them are not. \\
Obviously

	\begin{equation}
	\sum_{j_1,..,j_N=0,1} n(j_1,j_2,..,j_N) = n_T.
	\end{equation}

	\noindent Let $n_i(j_1,j_2,..,j_N)$ be the number of outcomes of a kind $i$ in the subset $(j_1,j_2,..,j_N)$. The assumption is that the events belonging to each subset $(j_1,j_2,..,j_N)$ are independent and identically distributed (i.i.d.). Such an assumption does not lead to loss of generality because the constraints used in building different subsets can contain the information about the dependencies between events \cite{GillespiePerspective}.\\
	\noindent Then if $n(j_1,j_2,..,j_N) \to +\infty$, the probability $\mathbb{P}_i(j_1,j_2,..,j_N)$ for the $i$-th outcome to occur in the set $(j_1,j_2,..,j_N)$ corresponds to the limit of frequency of the $i$-th outcome, i.e.:
	
	\begin{equation}
	\mathbb{P}_i(j_1,j_2,..,j_N) = \frac{n_i(j_1,j_2,..,j_N)}{n(j_1,j_2,..,j_N)} \quad \forall i \in \{1,..,N\}.
	\label{eqn:prob_1}
	\end{equation}

	\noindent A recent formulation of the Gillespie Stochastic Simulation Algorithm (SSA) \cite{GillespieReviewSSAminimumTime} can be used to define the probability $\mathbb{P}_i(j_1,j_2,..,j_N)$. Let   independent random variables $T_1, T_2, .. , T_N$ be the times required for the next occurrence of the respective outcome. The SSA suggests to pick the outcome that realizes the minimal occurrence time among the possible ones. In other words,  
	
	\begin{equation}
	\mathbb{P}_i(j_1,j_2,..,j_N) = 
	\begin{cases}
	0 & \mbox{if } j_i=0\\
	\mathbb{P}\left(T_i <  T_k: \, \forall k \neq i \mbox{ s.t. } j_k=1 \right)
	& \mbox{if } j_i=1 \, \land \, \exists k \neq i \mbox{ s.t. } j_k=1\\	
	1 & \mbox{if } j_i=1 \, \land \, j_k=0 \, \forall k \neq i
	\end{cases}.
	\label{eqn:prob_Gill}
	\end{equation}
	
	\noindent  Equations \eqref{eqn:prob_1} and \eqref{eqn:prob_Gill} offer the way for calculation of the asymptotic numbers of outcomes of a kind $i$, $n_i$, for any $i$:

	\begin{equation}
	n_i = \sum_{j_1,..,j_N} \mathbb{P}_i(j_1,j_2,..,j_N) n(j_1,j_2,..,j_N)\quad \forall i \in \{1,..,N\}
	\label{eqn:compute_general_ni}.
	\end{equation}
	
	\noindent The ratio between $n_i$ and $n_k$ for any $i$ and $k$ can be immediately obtained from   \eqref{eqn:compute_general_ni} and \eqref{eqn:prob_Gill}. The number of possible outcomes $N$ should not be crucial for the proposed approach. In particular, the proposed solution holds for big values of $N$. More important for the applicability of the method are the possible dependencies between different events. In general, complex dependencies between different subsets may lead to non-converged asymptotic behaviour of the ratios of interest. However, when the constraints are limited and well defined, as in the examples presented in the following sections, the ratios of interest can be evaluated exactly for any value of $N$.		
		
\section{A Process with a Single Constraint}	
\label{sec:one_constraint}
		
	\noindent Before applying the suggested methodology to a fitting procedure outlined in \autoref{sec:intro} we test it on the simple model introduced in \cite{TimeDelayedPDF}. We consider the case of a stochastic process with only two possible outcomes, 1 and 2. The first one is free to occur with the occurrence rate $c_1$, but the second one must wait till at least $n_0$ occurrences of kind 1 after its own previous occurrence. Its occurrence rate is $c_2$. 
	\noindent There are only two possible subsets in this case (the order of events is preserved as described above):
	
	\begin{equation}
	(1,0), \, (1, 1).
	\end{equation}
	
	\noindent The corresponding probabilities are given by

	\begin{align}
	\mathbb{P}_1(1,0) = 1, \quad & 
	\mathbb{P}_1(1,1) = \mathbb{P}(T_1<T_2), 
	\nonumber \\ 
	\mathbb{P}_2(1,0) = 0, \quad 
	& \mathbb{P}_2(1,1) = \mathbb{P}(T_2<T_1).
	\end{align}
	
	\noindent The total number of events $n(1,0)$ in the subset $(1,0)$ is given by $n_0$ outcomes 1 for each occurrence of the outcome 2:
	
	\begin{equation}
	n(1,0) = n_2 n_0,
	\end{equation}

	\noindent where $n_2$ is the asymptotic total number of outcomes 2. The total number of events in the complementary subset $n(1,1)$ can be computed by subtraction:

	\begin{equation}
	n(1,1) = n_T - n_2 n_0.
	\end{equation}

	\noindent \autoref{eqn:compute_general_ni} gives
	
	\begin{equation}
	n_1 = n_2 n_0 + \mathbb{P}(T_1<T_2) (n_T - n_2 n_0),
	\label{eqn:balance_n_1}
	\end{equation}
	
	\begin{equation}
	n_2 = \mathbb{P}(T_2<T_1) (n_T - n_2 n_0),
	\label{eqn:balance_n_2}	
	\end{equation}
	
	\noindent where $T_1$ and $T_2$ are the times required to fire the next occurrences of kind 1 and 2 respectively.\\
	\noindent Equations \eqref{eqn:balance_n_1} and \eqref{eqn:balance_n_2} can be rewritten as 
	
	\begin{equation}
	n_1 = \frac{ \mathbb{P}(T_1<T_2) + n_0 \mathbb{P}(T_2<T_1) } {1 + n_0 \mathbb{P}(T_2<T_1) } n_T,
	\label{eqn:n_1}
	\end{equation}
	
	\begin{equation}
	n_2 = \frac{ \mathbb{P}(T_2<T_1) } {1 + n_0 \mathbb{P}(T_2<T_1) } n_T.
	\label{eqn:n_2}	
	\end{equation}
	
	\noindent Equations \eqref{eqn:n_1} and \eqref{eqn:n_2} show that as $n_T \to \infty$, the asymptotic fraction $\nicefrac{n_2}{n_1}$ reaches the fixed asymptotic value. In particular
	
	\begin{equation}
	\frac{n_2}{n_1} = \frac{ \mathbb{P}(T_2<T_1) }{ \mathbb{P}(T_1<T_2) + n_0 \mathbb{P}(T_2<T_1) }.
	\label{eqn:n_ratio}	
	\end{equation}
	
	\noindent Following \cite{GillespieReviewSSAminimumTime}, we assign independent exponentially distributed probability density functions to the random variables $T_1$ and $T_2$: 

	\begin{equation}
	T_1 \ind \textrm{Exp}(c_1), \quad T_2 \ind \textrm{Exp}(c_2).
	\end{equation}
	
	\noindent The probabilities in \eqref{eqn:n_ratio} can be computed as suggested in \autoref{app:probabilities},  
	which gives
	\begin{equation}
	\mathbb{P}(T_1<T_2) = \frac{c_1}{c_1+c_2}, \quad \mathbb{P}(T_2<T_1) = \frac{c_2}{c_1+c_2}.
	\end{equation}
	
	\noindent Thus, the asymptotic ratio between the number of outcomes 2 and the number of outcomes 1 becomes
	
	\begin{equation}
	\frac{n_2}{n_1} = \frac{ c_2 }{ c_1 + n_0 c_2 }.
	\label{eqn:solution_single_constraint}
	\end{equation}
	
\begin{figure}[t]
    \centering
    \includegraphics[scale=0.6]{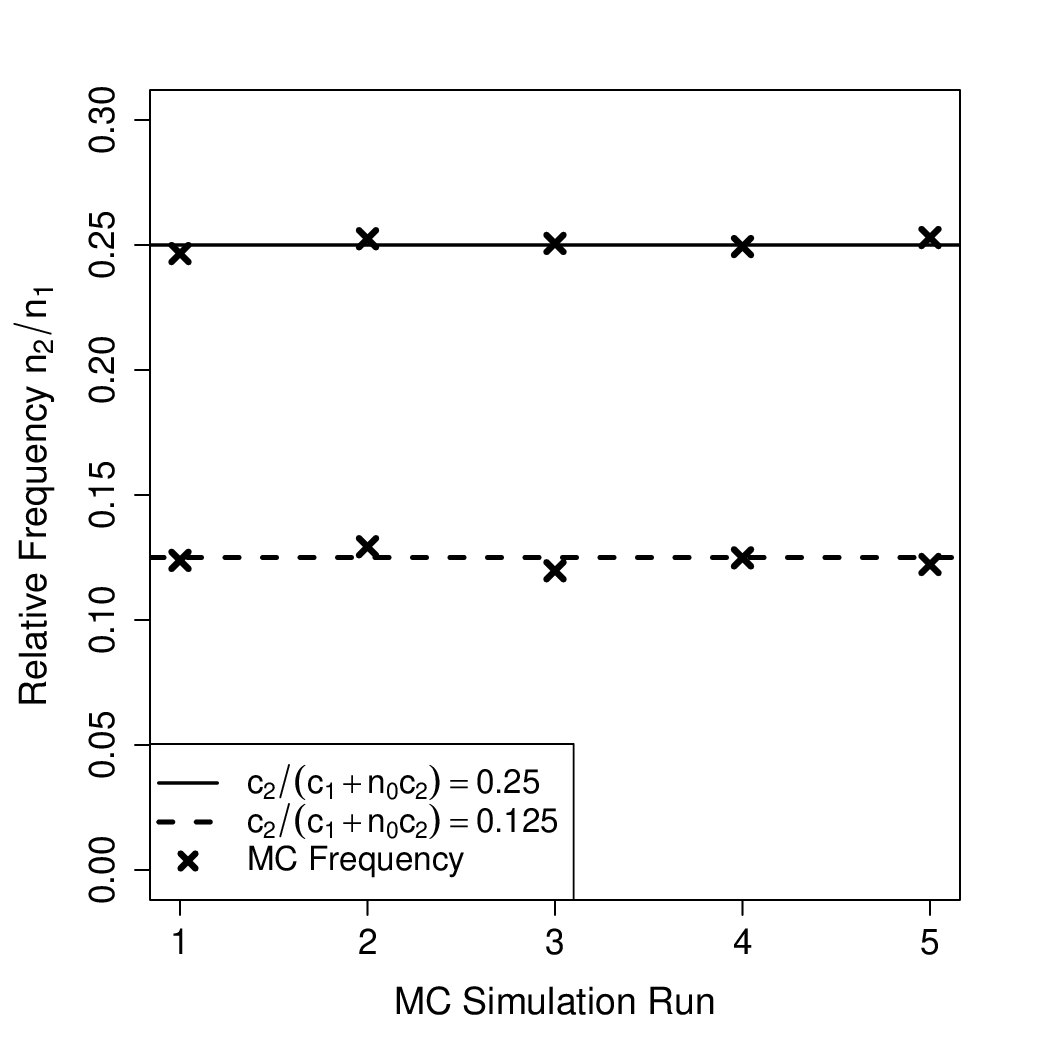}
    \caption{Comparison between the analytical solution \eqref{eqn:solution_single_constraint} (lines) and corresponding statistics (crosses) obtained by the Monte Carlo (MC) method proposed in \cite{TimeDelayedPDF}. Five independent runs are performed for two different parameters set: $n_0=3$, $\nicefrac {c_2}{c_1}=1$ (solid line) and $n_0=3$, $\nicefrac {c_2}{c_1}=0.2$ (dashed line). The MC sample size is equal to $G=10^4$.}
    \label{fig:single_constr}
    \end{figure}

	\noindent This result confirms the validity of the analytical approach presented for this model in \cite{TimeDelayedPDF}. It is also in good agreement with the data obtained using the Monte Carlo method proposed in \cite{TimeDelayedPDF}, as is shown in \autoref{fig:single_constr}.
	 
\section{Controlled Radical Polymerization}
\label{sec:CRP}

	\noindent In this section we introduce a real phenomenon that can be studied by the proposed analytical method. In particular, we are interested in a polymerization process known as Controlled Radical Polymerization (CRP) of acrylic monomers \cite{CRPexplaned}.\\
	
\begin{figure}[h]
    \centering
    \includegraphics[scale=0.6]{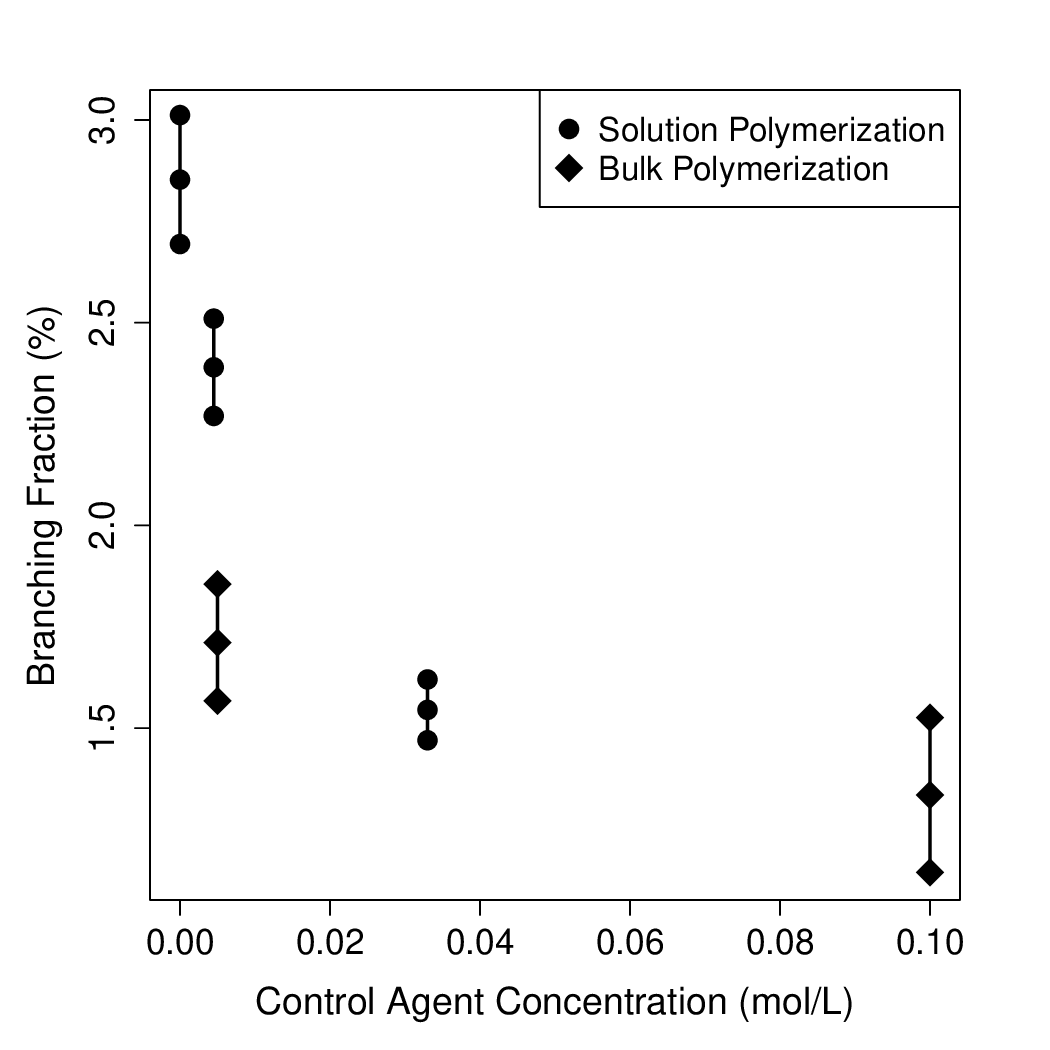}
    \caption{Experimental branching fractions and corresponding uncertainty intervals \cite{LinExpPdf}.}
    \label{fig:exp_data}
    \end{figure}

	\noindent Radical polymerization is a method by which high molecular weight polymer molecules can be formed by successive addition of individual monomer units. In radical polymerization of vinyl monomers, each addition, or \textit{propagation}, outcome regenerates the active radical species at the chain end, and the chain continues to grow until it terminates, typically by reaction with another radical.\\
	\noindent Alternatively, the reactive radical at the end of the chain can be transferred to a carbon atom within the chain, thus generating a so called \textit{mid-chain} radical. In radical polymerization of acrylic monomers this is particularly important and it occurs via a process (outcome) known as \textit{backbiting}. Due to the specific molecular orientation required for the backbiting outcome to occur, at least three propagation outcomes must take place before the next backbiting outcome occurs.\\
	\noindent Addition of monomer units to the mid-chain radical, which is formed by backbiting, results in the formation of a branched structure. Obtaining information about the kinetics of the process,  and the relative rate of reactions between backbiting and propagation, are important for understanding polymerization of acrylic monomers.  The ratio of the two competitive outcomes has a strong impact on polymer micro-structure and the mechanical properties of the resulting polymer. It can be measured by evaluating the branching fraction, which is determined experimentally as the ratio of the number of branches compared to the number of propagation outcomes that have occurred.\\
	\noindent Controlled Radical Polymerization (CRP) is a special type of radical polymerization which is conducted in the presence of an additional chemical reagent known as a control agent. In CRP the reactive radical chain end is subjected to frequent deactivation and reactivation steps allowing to control molecular weight of the polymer chain. Reversible deactivation of the reactive chain end is an outcome which occurs in competition with propagation and backbiting. We refer to it as the \textit{deactivation} outcome.\\
	\noindent Although classical chemical reaction kinetics dictate that the imposition of the additional competitive process of chain deactivation should not impact on the ratio of backbitings to propagations, experimental evidence has shown that there is a strong reduction in branching fraction under CRP conditions \cite{Ahmad_et_al, LinExpPdf, ExperimentalLinExpPdf}. In order to explain this reduction the existence of a non-exponential probability density function has been proposed \cite{LinExpPdf}. In \cite{LinExpPdf}, a Monte Carlo fitting scheme was suggested for evaluation of the appropriate parameters of the non-exponential pdf in order to fit the experimental branching fractions.  The data provided in \cite{LinExpPdf}, shown in \autoref{fig:exp_data}, will be used for comparing our analytical approach with the MC based procedure. Two sets of data with corresponding uncertainty intervals were obtained by two different experimental procedures, known as \textit{bulk} and \textit{solution} polymerization. Each polymerization was conducted at different control agent concentrations, thus giving a range of data points for the fitting procedure. 	
    
\subsection{The Monte Carlo Fitting Scheme}    
\label{sec:MCfitting}

	The Monte Carlo Fitting Scheme for evaluation of the branching fractions \cite{LinExpPdf} combines a MC method with an optimization algorithm.
	
	\vspace{0.1in}
	
	\noindent The MC method follows the SSA \cite{GillespieReviewSSAminimumTime}, where independent linear exponential pdf,  $\textrm{Linexp}(b_i,\tau_i)$, are assumed for required times for each outcome. It means that the time required for a propagation is $T_p \ind  \textrm{Linexp}(b_p, \tau_p)$, for a backbiting is $T_r \ind  \textrm{Linexp}(b_r, \tau_r)$ and for a deactivation is $T_d \ind  \textrm{Linexp}(b_d, \tau_d)$. The linear exponential pdf is defined as:
	
	\begin{equation}
    T_i \sim  \textrm{Linexp}(b_i,\tau_i)
	\Leftrightarrow    
    f_{LE}(t;b_i,\tau_i) =
    \begin{cases}
    \frac{2}{b_i^2+2b_i\tau_i} t & \mbox{if } 0 \le t < b_i \\
    \frac{2b_i}{b_i^2+2b_i\tau_i} e^{-\left(\frac{t-b_i}{\tau_i}\right)} & \mbox{if } t \ge b_i
    \end{cases}.
    \label{eqn:expressionlinexppdf}
    \end{equation}

	\noindent Different concentrations of monomers and control agents are considered as it is proposed in the supporting information of \cite{LinExpPdf}. The MC method is summarized in Algorithm \autoref{algo:MCmethod}.
		
	\vspace{0.1in}	
	
	\noindent The optimization routine can be described as follows: 
	
	\begin{enumerate}
    \item \label{item:point1} Choose the values of the parameters $\{ b_p, \tau_p, b_r, \tau_r, b_d, \tau_d \}$ for a particular attempt of the chosen optimization algorithm. 
    \item \label{item:compute_r} Run the MC method proposed in Algorithm \autoref{algo:MCmethod} and calculate the mean ratio between numbers of backbitings and propagations, $r$, corresponding to the current choice of parameters.
    \item Choose and compute the cost function $J$. The middle point of each uncertainty interval is used.
    \item Restart from \autoref{item:point1} until $J$ is minimized.
    \end{enumerate}	
    
    \noindent We have to stress that for the purposes of this paper the choice of an optimization algorithm is not important. However, in practice one should go for the most efficient one available, since this would help to reduce the number of Monte Carlo runs required for finding the optimal parameters. Next we show how the MC algorithm in the optimization scheme described above can be replaced with a significantly more effective analytical approach.\\
    \noindent The reason to invoke Monte Carlo procedure is because it is one of the most commonly used method to study such processes. Thus, we began by defining the kinetics of the system in terms of Monte Carlo rules, or by defining the explicitly constrained pdf proposed in \autoref{algo:MCmethod}. Next we show how to directly solve the dynamics of the system without having to resort to MC approach.
	
	\begin{algorithm}[!h]	
	Initialize the sample size $G$ and the index $i=0$\; 
	Initialize the number of occurred propagations $n_p=0$\;
	Initialize the number of occurred backbitings $n_r=0$\;
	Initialize the number of occurred deactivations $n_d=0$\;
	Initialize the counter for previous propagations $c_p=0$\;
	Assign the number of required propagations to have a backbiting $n_0=3$\; 		
	\While{ $i < G$  }{
		\eIf{ $ c_p < n_0 $ }{
			Draw an independent realization $t_p$ from $T_p \sim \textrm{Linexp}(b_p, \tau_p)$\;
			Draw an independent realization $t_d$ from $T_d \sim \textrm{Linexp}(b_d, \tau_d)$\;		
			\eIf{$ t_p = \min\{t_p,t_d\} $}{ 
				Propagation has occurred: $ n_p \gets n_p + 1 $\; 
				Update $ c_p \gets c_p +1 $\;
			}{ Deactivation has occurred: $ n_d \gets n_d + 1 $\; }			
		}{	Draw an independent realization $t_p$ from $T_p \sim \textrm{Linexp}(b_p, \tau_p)$\;
			Draw an independent realization $t_r$ from $T_r \sim \textrm{Linexp}(b_r, \tau_r)$\;
			Draw an independent realization $t_d$ from $T_d \sim \textrm{Linexp}(b_d, \tau_d)$\;
			\If{$ t_p = \min\{t_p,t_r,t_d\} $}{ 
				Propagation has occurred: $ n_p \gets n_p + 1 $\; }		
			\If{$ t_r = \min\{t_p,t_r,t_d\} $}{ 
				Backbiting has occurred: $ n_r \gets n_r + 1 $\;
				Reset the counter of previous propagations $c_p=0$\; }	
			\If{$ t_d = \min\{t_p,t_r,t_d\} $}{ 
				Deactivation has occurred: $ n_d \gets n_d + 1 $\; } }	
		Update $i \gets i+1$\; }
	Return the value of the branching fraction $r = \nicefrac{n_r}{n_p}$\;	
    \caption{The MC method for simulation of the evolution of Controlled Radical Polymerization carried out in the presence of control agent. The method returns the final branching fraction of the created chain.}  
    \label{algo:MCmethod}
    \end{algorithm}
      
\newpage  
    
\subsection{Analytical Expressions for Asymptotic Relative Frequencies of Events in CRP}
\label{sec:anal_sol}

	We will follow the ideas described in \autoref{sec:generic_asymptotic_fractions}. In particular, let $n_p$, $n_d$ and $n_r$ respectively be the asymptotic mean number of occurred propagations, deactivations and backbitings. Then the branching fraction can be calculated as a ratio between $n_r$ and $n_p$ using equations \eqref{eqn:prob_Gill} and \eqref{eqn:compute_general_ni}. As stated in \autoref{sec:CRP}, propagation and deactivation are always possible, whereas backbiting needs at least $n_0=3$ previous propagations to occur. Hence, we have two subsets (the order is: propagation
(p) , deactivation (d) , backbiting (r)):
	
	\begin{equation}
	(1,1,0), \, (1, 1,1).
	\end{equation}
	
	\noindent The following are the probabilities for each outcome in each subset:
	
	\begin{align}
	\mathbb{P}_p(1,1,0) = \mathbb{P}(T_p<T_d), \quad & 
	\mathbb{P}_p(1,1,1) = \mathbb{P}(T_p<T_d,T_r),
	\nonumber \\
	\mathbb{P}_d(1,1,0) = \mathbb{P}(T_d<T_p), \quad &
	\mathbb{P}_d(1,1,1) = \mathbb{P}(T_d<T_p,T_r),
	\nonumber \\
	\mathbb{P}_r(1,1,0) = 0, \quad &	
	\mathbb{P}_r(1,1,1) = \mathbb{P}(T_r<T_p,T_d).
	\end{align}
	
	\noindent The total number of events that cannot be a backbiting, $n(1,1,0)$, is given by $n_0$ propagations for each occurred backbiting, and by the number of deactivations occurred when backbiting is not possible:
	
	\begin{equation}
	n(1,1,0) = n_0 n_r + n_d(1,1,0 ) = 
	n_0 n_r + \mathbb{P}(T_d<T_p) n(1,1,0).
	\label{eqn:balance_not_back_ev}
	\end{equation}
	
	\noindent \autoref{eqn:balance_not_back_ev} can be rewritten as
	
	\begin{equation}
	n(1,1,0) = \frac{ n_0 n_r } { \mathbb{P}(T_p<T_d) }.
	\label{eqn:not_back_ev}
	\end{equation}
	
	\noindent Apparently,
	
	\begin{equation}
	n(1,1,1) = n_T - n(1,1,0).
	\end{equation}

	\noindent Then, \autoref{eqn:compute_general_ni} yields
	
	\begin{equation}
	n_p = \mathbb{P}(T_p<T_d) n(1,1,0)  + \mathbb{P}(T_p<T_d,T_r) [ n_T - n(1,1,0) ],
	\label{eqn:n_p}
	\end{equation}	
	
	\begin{equation}
	n_d = \mathbb{P}(T_d<T_p) n(1,1,0)  + \mathbb{P}(T_d<T_p,T_r) [ n_T - n(1,1,0) ],
	\label{eqn:n_f}
	\end{equation}	
	
	\begin{equation}
	n_r = \mathbb{P}(T_r<T_p,T_d) [ n_T - n(1,1,0) ].
	\label{eqn:n_b}
	\end{equation}
		
	\noindent From Eqs. \eqref{eqn:n_p} and \eqref{eqn:n_b} we have:
	
	\begin{equation}
	n_r = 
	\frac{ \mathbb{P}(T_p<T_d) \mathbb{P}(T_r<T_p,T_d) }
	{ \mathbb{P}(T_p<T_d) + n_0\mathbb{P}(T_r<T_p,T_d) } n_T,
	\label{eqn:n_b_solved}
	\end{equation}
		  
	\begin{equation}
	n_p =
	\frac{ \mathbb{P}(T_p<T_d) [ \mathbb{P}(T_p<T_d,T_r) + n_0 \mathbb{P}(T_r<T_p,T_d)] }
	{ \mathbb{P}(T_p<T_d) + n_0\mathbb{P}(T_r<T_p,T_d) } n_T.	
    \label{eqn:n_p_solved}
    \end{equation}
    
    \noindent Equations \eqref{eqn:n_b_solved} and \eqref{eqn:n_p_solved} show that as $n_T \to \infty$, the branching fraction $r=\nicefrac{n_r}{n_p}$ reaches the fixed asymptotic value. In particular, the ratio $r$ is a function of probabilities  involving the random variables $T_p$, $T_d$ and $T_r$ only.\\
    \noindent This implies two important consequences: (i) our solution holds for any choice of inter-events times pdf and (ii) it is possible to express the branching fraction as a function of the pdf parameters, once the probabilities of interest are computed. We explain how to compute the required probabilities for a generic case in \autoref{app:probabilities}. This method is still valid for the linear exponential case \eqref{eqn:expressionlinexppdf} considered in this study.\\
    
    \noindent The proposed analytical approach is based on the assumption that the asymptotic limit is reached. Thus, it  describes the statistics of the asymptotic states of  a stochastic process. Reaching the asymptotic
limit via Gillespie type of simulations is costly, since also the transient dynamics need to be performed. Availability of analytical shortcuts avoids having to deal with fluctuations present in finite systems. 
In particular, in the considered application (CRP), the reaction continues until all the reactants are used up. For this reason, the our method agrees with the experimental data for a reaction which may be considered to have reached its asymptotic state.     

\subsection{Numerical Results}
\label{sec:results}

\begin{figure}[!hp]
	\centering
  	\includegraphics[scale=0.6]{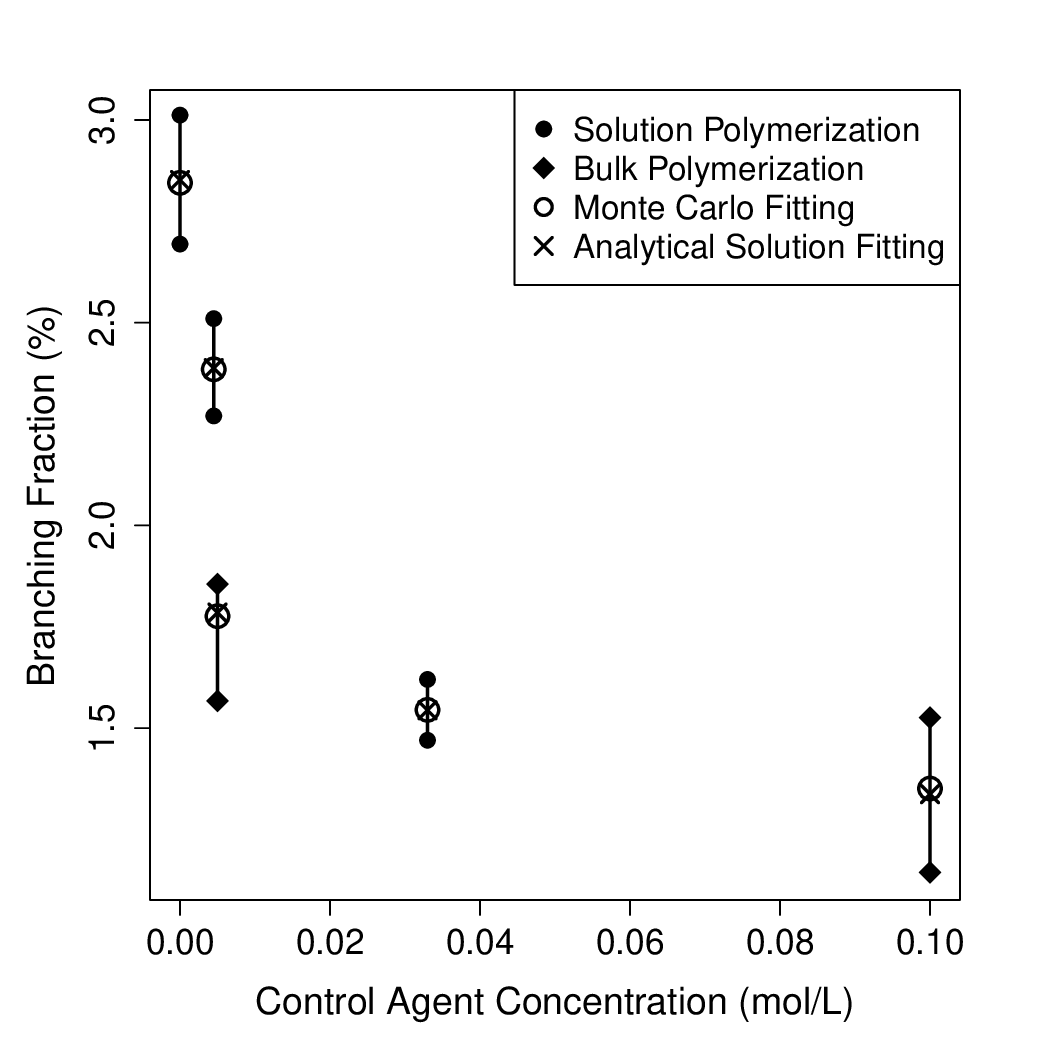}
  	\caption{Fitted data obtained by the analytical approach (crosses) and by the MC method (open circles) (MC sample size $G=10^4$) are presented for two polymerization reactions, bulk and solution. Both approaches use the linear exponential inter-events times pdf \eqref{eqn:expressionlinexppdf}.}
  	\label{fig:fitting_NM_LE}
	\end{figure}

\begin{figure}[!hp]
  	\centering
 	\includegraphics[scale=0.6]{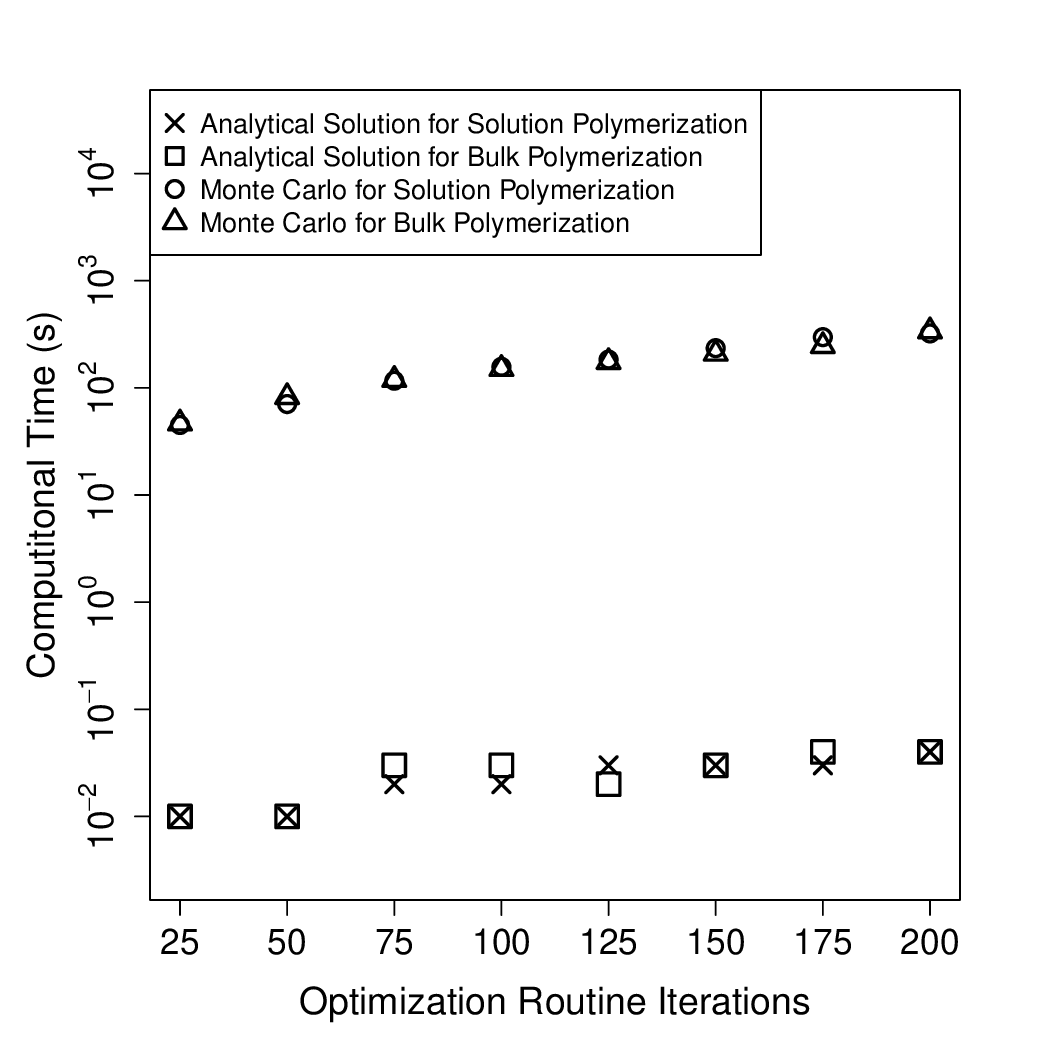}
 	\caption{Computational times required for the optimization routine (Nelder-Mead method \cite{NelderMeadmethod}) performed with an increasing number of iterations for bulk and solution polymerization. The analytical approach (crosses and squares) speeds up the procedure by the factor of $10^4$ compared with the MC based optimization method (open circles and triangles) of the same level of accuracy.}
  	\label{fig:CPUtime_NM_LE}
	\end{figure}

\begin{figure}[!hp]
	\centering
  	\includegraphics[scale=0.6]{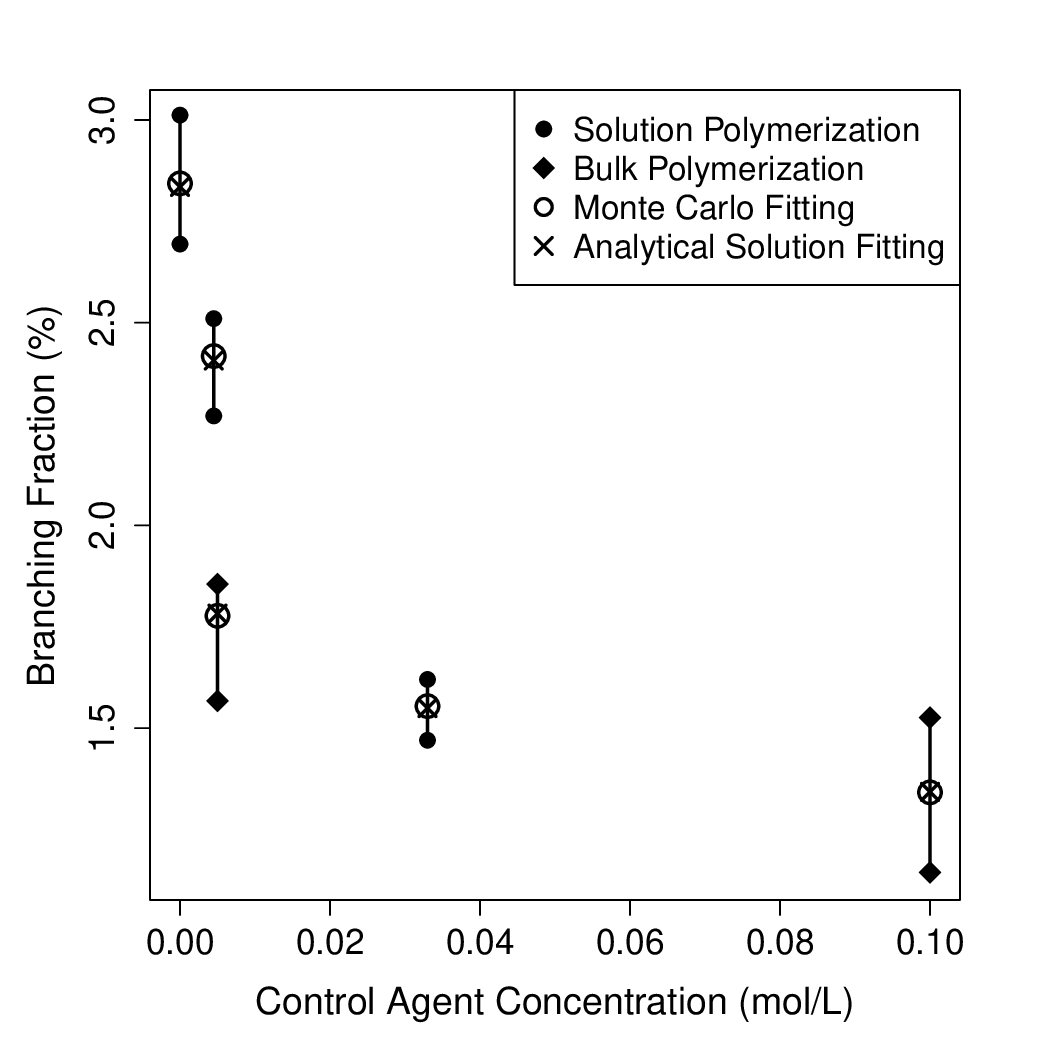}
  	\caption{Fitted data obtained by the analytical approach (crosses) and by the MC method (open circles) (MC sample size $G=10^4$) are presented for two polymerization reactions, bulk and solution. Both approaches use exponential pdf's for propagation and deactivation, and a linear exponential pdf \eqref{eqn:expressionlinexppdf} for backbiting.}
  	\label{fig:fitting_GEN_DSM}
  	\end{figure}

\begin{figure}[!hp]
  	\centering
 	\includegraphics[scale=0.6]{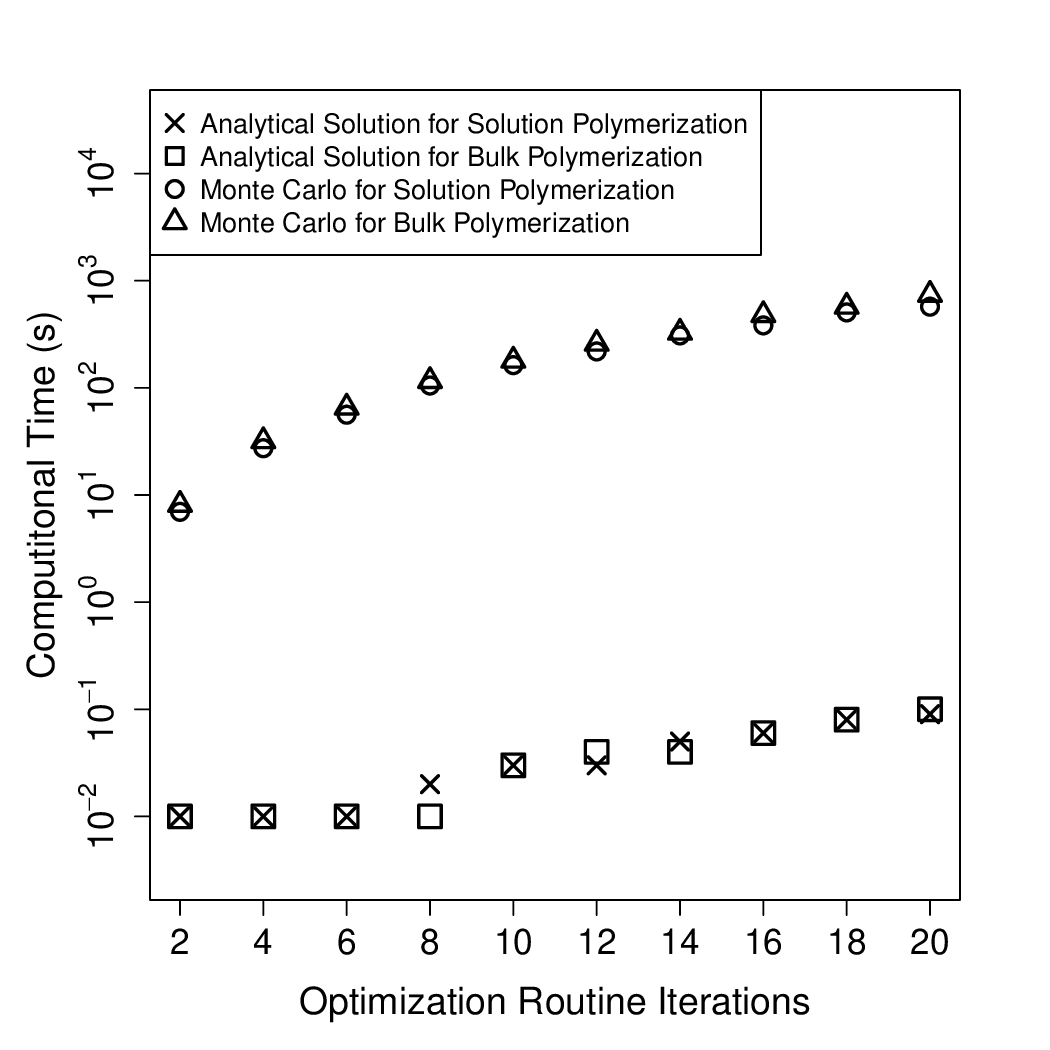}
 	\caption{Computational times required for the optimization routine (Genetic Algorithm \cite{GeneticOpt}) performed with an increasing number of iterations for bulk and solution polymerization. The analytical approach (crosses and squares) speeds up the procedure by the factor of $10^4$ compared with the MC based optimization method (open circles and triangles) of the same level of accuracy.}
	\label{fig:CPUtime_GEN_DSM}
	\end{figure}

	\noindent We present the results of numerical tests run for validation of the proposed approach.  The experimental data used in the tests were described in \autoref{sec:CRP}.\\	
	\noindent In \autoref{fig:fitting_NM_LE}, we assess the accuracy provided by two fitting approaches for calculation of the branching fractions using the experimental observations. The first approach (the data are shown by open circles) corresponds to the optimization routine explained in \autoref{sec:MCfitting}. The second method (crosses) follows the same optimization route (Nelder-Mead method \cite{NelderMeadmethod}) and evaluates the same cost function $J$, but the Monte Carlo evaluation of the branching fractions is replaced with the analytical expressions given in \eqref{eqn:n_b_solved}, \eqref{eqn:n_p_solved}. Both, the MC method and the analytical approach, use a linear exponential inter-events time pdf \eqref{eqn:expressionlinexppdf}. As it follows from \autoref{fig:fitting_NM_LE}, the sample size $G=10^4$ in the Monte Carlo approach  guarantees the same level of accuracy provided by the analytical method. \autoref{tab:optimized_param_LE} shows the optimized parameters for the linear exponential inter-events time pdf \eqref{eqn:expressionlinexppdf}.\\
	\noindent Although both methods can offer comparable accuracies, it is not the case for the computational cost. The optimization routine performed with the analytical approach is up to $10^4$ times faster than the one using the MC method of the same level of accuracy (MC sample size $G=10^4$). Comparative computational times are shown in \autoref{fig:CPUtime_NM_LE}.\\
	\noindent It is clear that the degree of speed up provided by the analytical approach over the MC method is determined by the MC sample size $G$. Indeed, the computational complexity of the analytical method is $O(1)$ whereas it is $O(G)$ in the case of Monte Carlo. This is confirmed by the numerical tests.\\		
	\noindent Different optimization algorithms and inter-events time pdf were tested. In \autoref{fig:fitting_GEN_DSM}, \autoref{fig:CPUtime_GEN_DSM} and \autoref{tab:optimized_param_DSM} we present the results obtained with the fitting scheme in which both the optimization routine and the inter-events time pdf differ from those applied in the previous test.  In particular, a Genetic Algorithm \cite{GeneticOpt} has been selected for optimization. Pure exponential pdf was assigned for propagation and deactivation, whereas linear exponential pdf \eqref{eqn:expressionlinexppdf} was chosen for backbiting. The rationale behind the choice of the pdf for propagation and deactivation is confirmed by our previous test, described above, which results in very small optimal parameters $b_p$ and $b_d$.  This suggests that the optimal pdf choices for propagation and deactivation are very close to an exponential function. The results shown in \autoref{fig:fitting_GEN_DSM} justify this choice. Also, these results confirm that the proposed analytical approach is valid for various choices of optimization routines and inter-events time pdf. 
	
\newpage
	
	\begin{table}[!hp]
	\centering
	\begin{tabular}{|c|c|c|c|c|c|c|c|}
    \hline
	Polymerization & Fitting & $b_p$ & $b_r$ & $b_d$ & $\tau_p$ & $\tau_r$ & $\tau_d$ \\
	\hline
	Solution & AS & $1.74 \times 10^{-1}$ & $6.53$ & $2.28 \times 10^{-4}$ & $ 9.1 \times 10^{-1}$ & $1.31$ & $3.58 \times 10^{-2}$\\
	\hline
	Solution & MC & $1.74 \times 10^{-1}$ & $6.53$ & $2.28 \times 10^{-4}$ & $ 9.1 \times 10^{-1}$ & $1.31$ & $3.58 \times 10^{-2}$\\
	\hline
	Bulk & AS & $2.8 \times 10^{-1}$ & $1.58  \times 10^{-1}$ & $1.57 \times 10^{-2}$ & $ 8.53 \times 10^{-1}$ & $11.54$ & $ 3.43 \times 10^{-2}$\\ 
	\hline
	Bulk & MC & $1.64 \times 10^{-1}$ & $1.40  \times 10^{-1}$ & $3.557 \times 10^{-2}$ & $ 9.16 \times 10^{-1}$ & $12.01$ & $ 2.56 \times 10^{-2}$\\ 
	\hline
    \end{tabular}
    \caption{Optimized parameters of linear exponential inter-events times pdf \eqref{eqn:expressionlinexppdf} obtained by analytical solution fitting (AS) and Monte Carlo fitting (MC). All results assume unitary concentrations of monomers and control agents (supporting information can be found in Ref. \cite{LinExpPdf}).}
    \label{tab:optimized_param_LE}
    \end{table}

\newpage	
	
\newpage	
		
  	\begin{table}[!hp]
	\centering
	\begin{tabular}{|c|c|c|c|c|c|c|}
    \hline
	Polymerization & $b_p$ & $b_r$ & $b_d$ & $\tau_p$ & $\tau_r$ & $\tau_d$ \\
	\hline
	Solution & - & $5.94$ & - & $1$ & $5.97$ & $ 4.78 \times 10^{-2} $ \\
	\hline
	Bulk & - & $3.82  \times 10^{-1}$ & - & $1$ & $8.80$ & $1.33$ \\ 
	\hline
    \end{tabular}
    \caption{Optimized parameters corresponding to the model with exponential pdf's for propagation and deactivation, and a linear exponential pdf \eqref{eqn:expressionlinexppdf} for backbiting. In the case of exponential pdf, the parameter $\tau$ expresses the mean value of the corresponding random variable. All results are for unitary concentrations of monomers and control agents (supporting information can be found in Ref.  \cite{LinExpPdf}). The shown set of parameters was found with the analytical solution fitting. Then, the same set was tested with the MC method, providing the data fitting shown in \autoref{fig:fitting_GEN_DSM}.}
    \label{tab:optimized_param_DSM}
    \end{table}
    
\newpage
  	
\newpage

\section{Conclusions}
\label{sec:sum}

Stochastic simulation algorithms produce multiple realizations of the full evolution of studied processes. In principle, this information allows a systematic study of all properties of the simulated system. 
The price for this detailed description includes, however, excessive computational time, needed to perform the Monte Carlo  simulations, as well as the presence of unavoidable statistical errors.
If the focus is on extracting particular quantities of interest, it may be possible to solve the problem analytically.
 The proposed analytical approach is one example of this strategy. 
 We have demonstrated superiority of this analytical approach, over a traditional Monte Carlo simulation in the computation of the branching fraction in Controlled Radical Polymerization. 
 The approach is free of statistical errors, and thus guarantees more accurate estimations,  than those provided by a Monte Carlo simulation. 
 In addition, the method is significantly (an order of the sample size) faster than the Monte Carlo approach. The performed tests show that the choice of optimization algorithm is not important and the analytical approach works with any choice of the inter-events time pdf. The proposed method is also general enough to be used in other applications, especially in those involving constrained stochastic processes which are difficult and often impossible to simulate using a conventional Statistical Simulation Algorithm. 
Finally, the proposed approach can be used as an efficient tool for finding the optimal set of parameters for inter-events time pdf, to be further utilized in detailed Monte Carlo simulations.

\section{Acknowledgements}
\label{sec:ack}

This research is supported by the Spanish Ministry of
Economy and Competitiveness MINECO: BCAM Severo
Ochoa accreditation SEV-2013-0323, grant SVP-2014-068451.
One of us (EA) acknowledges the support of the project MTM2013-46553-C3-1-P. 
DS acknowledges support of the Basque Government (Grant No. IT-472-10), and of the Ministry of Science and Innovation of Spain (Grant No. FIS2009-12773-C02-01).	
The authors thank J.M. Asua (POLYMAT, Spain) for the valuable discussions.  
\appendix

\section{Computation of the Probabilities}
\label{app:probabilities}

	Below we explain how to compute the probability $\mathbb{P}(T_k<T_j \, \forall j \neq k)$, with $k,j \in \{1,..,N\}$. $T_1,..,T_N$ are independent random variables with the generic pdf $f_i(t)$, $i \in \{1,..N\}$:
	
	\begin{equation}
	T_i \ind f_i(t) \quad \forall i \in \{1,..,N\}.  
	\label{eqn:Ti_pdf}	
	\end{equation}  
	
	\noindent The cumulative density function $F_i(t)$ of the random variable $T_i$ is the following:
	
	\begin{equation}
	F_i(t) := \mathbb{P}(T_i \le t) = \int_0^{t} f_i(\tau) d\tau
	\quad \forall i \in \{1,..,N\}. 
	\label{eqn:Ti_cdf}	
	\end{equation}  
				
	\noindent The probability of interest is then computed as: 	
	\begin{align}
	\mathbb{P}(T_k<T_j \, \forall j \neq k) & 
	= \int_0^{+\infty} \mathbb{P}(T_k<T_j \, \forall j \neq k, \, T_k=\tau) d\tau =
	\nonumber \\
	& = \int_0^{+\infty} \mathbb{P}(T_k=\tau) \prod_{j \neq k} \mathbb{P}(T_j>\tau) d\tau =
	\nonumber \\		
	& = \int_0^{+\infty} f_k(\tau) \prod_{j \neq k} \left( 1-F_j(\tau) \right)  d\tau \quad \forall k \in \{1,..,N\}. 
	\label{eqn:compute_min_pb}
	\end{align}


\newpage	

\begin{thebibliography}{16}

\bibitem{GillespieOriginalSSA}
D.~Gillespie.
\newblock Exact stochastic simulation of coupled chemical reactions.
\newblock {\em Journal of {P}hysical {C}hemistry}, 81(25):2340--2361, 1977.

\bibitem{Bortz1975}
A.B. Bortz, M.H. Kalos, and J. L. Lebowitz.
\newblock A new algorithm for Monte Carlo simulation of Ising spin systems.
\newblock {\em Journal of Computational Physics}, 17(1):10--18, 1975.

\bibitem{GillespieReviewSSAminimumTime}
D.~Gillespie.
\newblock Stochastic {S}imulation of {C}hemical {K}inetics.
\newblock {\em Annu. {R}ev. {P}hys. {C}hem.}, 58:35--55, 2007.

\bibitem{ConstrainedSystemExamples}
Y.T. Yeh, L. Yang, M. Watson, N.D. Goodman, and Pat Hanrahan.
\newblock Synthesizing open worlds with constraints using locally annealed reversible jump MCMC.
\newblock {\em Journal ACM Transactions on Graphics (TOG)}, 31(4):4--56, 2012.

\bibitem{Ahmad_et_al}
N.M. Ahmad, B.~Charleux, C.~Farcet, C.J. Ferguson, S.G. Gaynor, B.S. Hawkett,
  F.~Heatley, B.~Klumperman, D.~Konkolewicz, P.A. Lovell, K.~Matyjaszewski, and
  R.~Venkatesh.
\newblock Chain {T}ransfer to {P}olymer and {B}ranching in {C}ontrolled
  {R}adical {P}olymerizations of n-{B}utyl {A}crylate.
\newblock {\em Macromol. Rapid Commun.}, 30:2002--2021, 2009.

\bibitem{ReyesArzamendi2011}
Y.~Reyes, G.~Arzamendi, J.M Asua, and J.R. Leiza.
\newblock {B}ranching at {H}igh {F}requency {P}ulsed {L}aser {P}olymerizations
  of {A}crylate {M}onomers.
\newblock {\em Macromolecules}, 44:3674--3679, 2011.

\bibitem{Barner-Kowollik2006}
C.~Barner-Kowollik, M.~Buback, B.~Charleux, M.L. Coote, M.~Drache, T.~Fukuda,
  A.~Goto, B.~Klumperman, A.B. Lowe, J.B. Mcleary, G.~Moad, M.J. Monteiro, R.D.
  Sanderson, M.P. Tonge, P.~Vana, and P.~Marie.
\newblock Mechanism and {K}inetics of {D}ithiobenzoate-{M}ediated {RAFT}
  {P}olymerization. {I}. {T}he {C}urrent {S}ituation.
\newblock {\em J. Polym. Sci., Part A: Polym. Chem.}, 44:5809--5831, 2006.

\bibitem{Moad}
G.~Moad.
\newblock Mechanism and {K}inetics of {D}ithiobenzoate-{M}ediated {RAFT}
  {P}olymerization - {S}tatus of the {D}ilemma.
\newblock {\em Macromol. Chem. Phys.}, 215:9--26, 2014.

\bibitem{Coelho}
J.F.J. Coelho, A.C. Fonseca, P.M.F.O. Gon{\c c}alves, A.V. Popov, and M.H. Gil.
\newblock Particle features and morphology of poly(vinyl chloride) prepared by
  living radical polymerisation in aqueous media. {I}nsight about particle
  formation mechanism.
\newblock {\em Polymer}, 52:2998--3010, 2011.

\bibitem{CRPexplaned}
G.~Moad and D.H. Solomon.
\newblock {\em The {C}hemistry of {R}adical {P}olymerization}.
\newblock {E}lsevier, ISBN 0-08-044286-2, {S}econd fully revised edition, 2006.

\bibitem{TimeDelayedPDF}
D.~Sokolovski, S.~Rusconi, E.~Akhmatskaya, and J.M. Asua.
\newblock Non-{M}arkovian models of the growth of a polymer chain.
\newblock {\em Proc. R. Soc. A}, 471: 20140899.

\bibitem{LinExpPdf}
N.~Ballard, S.~Rusconi, E.~Akhmatskaya, D.~Sokolovski, J.~de~la Cal, and J.M.
  Asua.
\newblock The {I}mpact of {C}ompetitive {P}rocesses on {C}ontrolled {R}adical
  {P}olymerization.
\newblock {\em Macromolecules}, 47(19):6580--6590, 2014.

\bibitem{GillespiePerspective}
D.T. Gillespie, A.~Hellander, and L.R. Petzold.
\newblock Perspective: {S}tochastic algorithms for chemical kinetics.
\newblock {\em J. {C}hem. {P}hys.}, 138:170901--1 -- 170901--14, 2013.

\bibitem{ExperimentalLinExpPdf}
N.~Ballard, M.~Salsamendi, J.I. Santos, F.~Ruip{\'e}rez, J.R. Leiza, and J.M.
  Asua.
\newblock Experimental {E}vidence {S}hedding {L}ight on the {O}rigin of the
  {R}eduction of {B}ranching of {A}crylates in {ATRP}.
\newblock {\em Macromolecules}, 47:964--972, 2014.

\bibitem{NelderMeadmethod}
J.A. Nelder and R.~Mead.
\newblock A simplex method for function minimization.
\newblock {\em Computer Journal}, 7(4):308--313, 1965.

\bibitem{GeneticOpt}
M.~Mitchell.
\newblock {\em An {I}ntroduction to {G}enetic {A}lgorithms}.
\newblock Cambridge, MA: MIT Press, 1996.

\end{thebibliography}
\end{document}